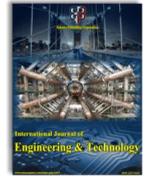

# Implementing DMZ in Improving Network Security of Web Testing in STMIK AKBA

Akbar Iskandar[1]*, Elisabet Virma[1], Ansari Saleh Ahmar[2]

[1]*Informatics Engineering, STMIK AKBA, Makassar, Indonesia*
[2]*Department of Statistics, Universitas Negeri Makassar, Makassar, Indonesia*
*\*Corresponding author E-mail: akbar.iskandar06@gmail.com*

**Abstract**

The aims of this research are to design and to implement network security system in internal web testing using DeMilitarized Zone Method and Microtic Router on Siakad server of STMIK AKBA. Data analysis techniques that possible to use is descriptive method. The significances of the study are 1) to avoid the attack of cracker who intend to access the system without permissions and 2) to improve network securityon web testing services on Siakad server of STMIK AKBA. The data are obtained by having literature review and observation. Literature review assists the researchers to collect the theory on DeMilitarised Zone Method and the previous studies which were used as comparison to the recent study. Obervation was carried out directly to the field to observe the running system. Based on the results and discussion, it is shown that the aplication of DeMilitarized Zone Method on microtic can secure the web testing on Siakad server of STMIK AKBA and can maintain the whole series of online services that are available in the server.

## 1. Introduction

Computer network security system is a way to prevent or protect and identify unauthorized users (intruders) to access a network in order to avoid various threats without any obstacles [1]. In addition, security system is also used to monitor and prevent unauthorized access, misuse of information, modification and even deletion of important data on a server. It is also stated by [2], [3] that the threats can be categorized into two types i.e. internal and external threats. Internal threat could exist intentionally or untentionally. External threat could be tapping system from others. The threat forms of computer network become varied day by day. Therefore the thing that should be kept in mind is no network of computers that is anti-tapping or no network is completely safe from various threats from people who have bad intention. Because the nature of network is having comunication openly and the information can be accessed also by people who intend to misuse them. Therefore, security system is highly needed to secure the data in a computer network. The precautions to protect the network is the duty of network administrator.

As it is known, the purpose of making network security is one way to anticipate the bad risk both physical and non-physical threats, either directly or indirectly because the level of crime in computer system could be classified from annoying to very dangerous ones [4]. Physical forms of crime that could lose information are the damage of computer and network communication devices, theft, computer hardware or network devices and natural disaster as well. In another side, non-physical category includes the damage of operation system, the damage of applications, the threat of virus and sniffing in which the crime is done by wiretapping techniques or through monitor to steal the personal data [5].

### 1.1. Computer Network Security

As it is known that computer network is a set of several computers which are connected each other in a network in order to have data communication between one computer to others. The form of data communication through computer network could be text, pictures, video, or voice. Network topology is a design or form that could describe how the computers are connected each other through computer network that connected together to share the data [6]. Each topology has its advantages and disadvantages.

Therefore, to establish a good computer network, both small and large scale, the first thing to do is designing the topology because the selection of network topolgy shall affect the network scale, the cost needed, communication model to be applied, the speed of access and the purpose of the network. This topology shall assist us to analyze the need of network hardware that will be used and the way to access each computer which is connected each other in that network.

Computer network technology or internet nowdays is often discussed in some sources that security system is very essential for a computer network system that connected to outside network or internet, because internet is a widely open computer network in the world. Therefore, the existance of device such as microtic could be used as network security system [7], [8].

Microtic router is one of operation system or software which could be used to make computer as a reliable network router, in which it has various features for network and wireless. In addition, microtic could be also act as a firewall for other computers and give priority for other computers in order to be able to access internet data and local data.

Firewall is a software that can allow the network traffic which is considered safe and can prevent the network traffic which is considered unsafe [9], [10]. Besides that, firewall is generally applied in machine running on a gateway between local network (LAN) and internet network which functions as security system [1].

Reference [9], [11], [2], [12] the reasons of the importance of network security:

- Privacy: The ability of individu or group to keep the personal data which is not allowed to access publicly.
- Confidentiality: Avoiding information to be accessed by unauthorized persons.
- Integrity: The obedence of the values or norms which are applied in one organization or professional code of ethics.
- Authentication: The process of identity proofing is to ensure that the users are true users and if it is proven that they are the owner of that identity, the relation will be done based on the given rights.
- Authorization: The process of determining what rights are given to the users who possess the clear idenntities.
- Availability: Related to the availability of information when it is needed.
- Access Control: As the central of computer security because the permit is given to the objcet to determine the persons who are able to access the object.
- Non-repudiation: An individual identification or device which is able to record the activities of users in order to make users not to deny the access that they have done.

Stand-alone computers have several limitations. Therefore, the existance of computer network will make computer able to do a lot of things and can make the efficiency and effectiveness of the information exchange and security system also becomes the major priority [13].

### 1.2. DeMilitarized Zone (DMZ)

One of approaches to tackle attacks from providers/attack target DDOS (*Distributed Daniel Of Service*) is the application of Client Puzzle protocol. This Client Puzzle Protocol aims to withstand from attacks that minimize the server capability to do service request in the beginning fase, i.e. connecting by using DeMilitarized Zone method which is managed in microtic device.

DMZ is the abbrivation of Demilitarized Zone which is well known as security layer and also as perimeter network that is used to protect internal system where all ports are open so that they are possibly seen by outsiders. Therefore, when there is attack or someone intentionally attack the server which uses DMZ, the attacker can only access host in DMZ, not in internal network [14]. The essential function of DMZ is to control the network traffic. It is because the basic working principle of DMS is to move all services of network from one network to another different network in order to avoid a single fail point that can cause violation of control system.

Therefore, the network is safe from intruders who intend to change the data or even delete the data in server. Computers in DMZ network are special computers which are able to contact directly by outsiders and the application used in each host on DMZ should be safe, and to be able to monitor or update regularly. For instance, on web server, mail exchangel server and name server [3], [7]. Futhermore, DMZ is established based on three concepts, i.e. 1) NAT (Network Address Translation) functions to show the packages from "real address" to internal address. 2) PAT (Port Addressable Translation) functons to show the data coming from particular port, or range of a port and protocol (TCP/UDP or others) and IP address to particular port or range of a port to internal IP address and Access List which functions to control precisely what entering and out from a network in question form.

Compared to the using of *client Puzzle, Demilitarized zone* is easier in accessing the server resources because it is not bothered by the puzzle. Another approach is using Load Balancing technique. Load Balancing is a technique to distribute the traffic loads into two or more connection paths in a balanced way to make the traffic run optimally, maximize the *throughput*, minimize response time and avoid over load in one of connection paths.

### 1.3. Web testing

Test, measurement, assesment and evaluation are inseparable entities because they are interconnected each other and have their own role and purpose in education [15]. Definition of test is a set of tasks or question or statement that have been planned to get information about one atribute. Furthermore, measurement is the process of scoring the test results that have been carried out or the result of measurement called score. In addition, assessment is a qualitative revealation that can be used to respond the result of measurement such as does the result of measurement good or bad? The last evaluation is recomendation or decision making [16].

Nowdays, conducting test, measurement, assessment and evaluatiaon has been done web-based in order to easily know the students capability level, to measure the students improvement, to know the curriculum accomplishment, to know the students learning difficulties.

Another example of application that can be used to measure the performance is web Science and Technology Index (SINTA). This application aims to measure the improvement performance of science and technology for institution, researchers about the researches that have been completed [17]. In addition, test application have been widely used in selection process of prospective students or employees more ever in process of promotion of employees. Those activities have utilized web engineering that is intergrated to computer network.

Currently, the system that had been applied in conventional way shifted to using technology although it has been known that technology has advantages and disadvantages. The advantages are to facilitate us in making and providing questions,setting the class, setting the users who are possible to take the test, setting the process of test to the reporting of test results [18]. Otherwise, the disadvantages are related to the network security.

There are some ways to measure the person ability both in written and oral. In order to prepare and conduct the test easily, the best way is applying web testing. In which, the web testing could be conducted by using computer and internet.

Applying the web testing often involves computer network then can be used to have online test without using papers (paperless) or now it is called Computer Based Test (CBT) or Computer Assisted Test (CAT) because all test activities using computers both in internet network and intranet network. Besides that, other advantages are its capability to give feedback to test takers and easy in administration [19].

In some universities, both public and private, the network security is always improved especially for the university as the objcet of this research, because the topology used now still utilizes system installing server or resources for local access and internet access within the same network group and connected directly to host server without using a firewall.Therefore, it should be evaluated.

Analyzing the condition of network security that needs improvement, the researcher designed and implemented the network security system in internal web testing by using DeMilitarized Zone and Microtic Router.

## 2. Methodology

The type of this research includes R & D (Research and Development) where the researchers conducted research and designed or even dveloped software and hardware that were then applied to the object of research.

The high number of attacks and cyber crimes to web application in private organization or public institutions in several countries is related to the negligence of objects that were attacked and because of the failure to detect vulnerabilityies which were zero-day. It is then easy to infiltrated by criminals who always prepare to ruffle the data in the system.

Therefore, the researchers design and implement the DeMilitarized Zoe and Microtic Router to secure the data of online test result using simulation access through Local Area Network (LAN) or internet. [20] stated that computers for server should be fast CPU, large hard drive, good RAM. Otherwise, the hardware needed to network implementation with DMZ method is:
Computer server with the minimum specification:
- Processor Intel dual core
- Hard Disk Drive 20 GB
- RAM / Memory 1024 MB
- Ethernet Card 100/1000 Mbps

Computer Client with the minimum specification:
- Processor Intel Celeron
- Hard Disk Drive 20 GB
- RAM / Memory 1024 MB
- Ethernet Card 100/1000 Mbps

Then, the software needed in designing the network security with DeMilitarized Zone method is:
In Computer client
- Operation system Windows 7
- Mozilla Firefox
- Nmap Scanner

In computer Server
- Operation system Linux ubuntu server 16.04
- Web server apache

In Microtic Device
- Microtic Router OS
- Firewall

### 2.1. Analysis of Weakness of Old System

The security condition of resources in the current research object should be improved because the topology still uses system that installs server for local access and internet access in the same network group and connects directly to host server without firewall. It may allow certain people to attack the resources easily. The simple scheme model used today before using DEMilitarized Zone Method can be seen in Figure 1.

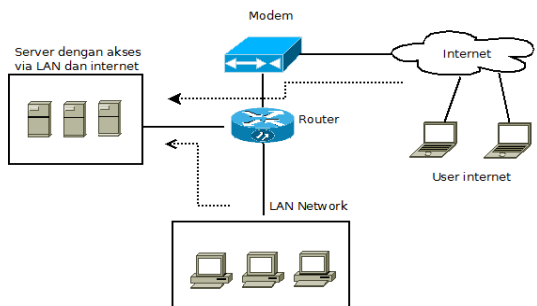

**Fig. 1:** Network Scheme without DMZ

### 2.2. Proposed System

The currently designed network security system seeks to establish a firewall zone of Demilitarized Zone *DMZ)* in STMIK AKBA by using simulation access through Local Area Network (LAN) or internet. DMZ Firewall or perimeter network is security network boundary located betwen corporate/private LAN network and public network. With the firewall, all traffics are forced to pass through one single concentrated check point where all traffics will be controlled.
Giving segmentation to firewell system is useful to protect server in LAN Corporate network from hacker attacks. The designing system of network security with DeMilitarized Zone Method is shown in Figure 2.

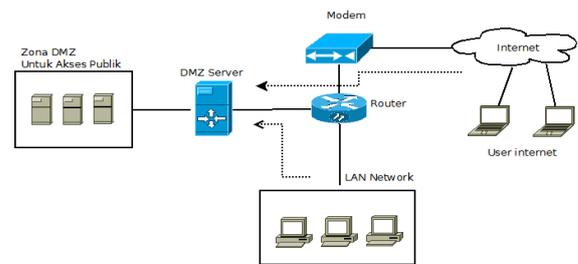

**Fig. 2:** Network Scheme using DMZ

The design of new system uses microtic device as DMZ server. Here, DMZ server will be a protector when someone tries to access the resources in server at DMZ zone from internet. As the trial in this simulation, user will access the resources which is web server running in a Operating System Open Source Linux Ubuntu 16.04 as Linux product and widely used by people. In default, system in web server will run at local network and accessed from port 80 but when accessing from internet, the local network will be translated to a public IP and port 80 will be directed to port 81.

### 2.3. Data Colection Method

Collecting the research data was carried out by doing literature study and observation. Collecting data by literature study was conducted to find references and theories related to the focus of research being done. It aims to have enough knowledge and theoretical basis that support to find solution of research problems related to network security system of web testing using DeMilitarised Zone Method.
Futhermore, in Literature study, researchers studied about previous related studies as comparison to the current study. Meanwhile observation is observing the object in the field and it can be done directly or indirectly in order to get information as a basis in conducting the research.

## 3. Findings and Discussion

### 3.1. The Results Analysis of old system

Based on the observation of old system to research object, the description of network topology was found as shown in Figure 3. In this Figure, it is shown that the existing server in STMIK AKBA have less secure access when accessing from outer network or internet. This is because the server doesn't have protection such as DMZ server. Therefore, to examine the network security system of server in STMIK AKBA ,the researchers scan the server by using NMAP Tool Scanner and can be seen in picture 3 below:

```
Starting Nmap 6.40 ( http://nmap.org ) at
2017-08-03 03:32 Malay Peninsula Standard Time
NSE: Loaded 110 scripts for scanning.
NSE: Script Pre-scanning.
Initiating Ping Scan at 03:32
Scanning siakad.akba.ac.id (          ) [4
ports]
Completed Ping Scan at 03:32, 0.29s elapsed (1
total hosts)
Initiating Parallel DNS resolution of 1 host.
at 03:32
Completed Parallel DNS resolution of 1 host.
at 03:33, 13.01s elapsed
Initiating SYN Stealth Scan at 03:33
Scanning siakad.akba.ac.id (          ) [1000
ports]
Discovered open port 8888/tcp on
Discovered open port 443/tcp on
Discovered open port 110/tcp on
Discovered open port 21/tcp on
Discovered open port 993/tcp on
....................
```
**Fig. 3:** The scan result using the NMAP scanner tool

After scanning as shown in Figure 3, the informations about open ports, IP server, and active services in computer server can be seen. It also shows the operating system using in web server in STMIK AKBA before applying DMZ method i.e. Apache version 2.4.7. for more details, the scan result can be seen in Figure 4.

```
3/tcp     open  tcpwrapped
|_auth-owners: ERROR: Script execution failed
(use -d to debug)
4/tcp     open  tcpwrapped
|_auth-owners: ERROR: Script execution failed
(use -d to debug)
......................
80/tcp    open  http Apache httpd 2.4.7
((Ubuntu))
|_auth-owners: ERROR: Script execution failed
(use -d to
 debug)
|_http-favicon: Unknown favicon MD5:
A2C8A7FCE771F22317A559D8D8ADD979
|_http-methods: No Allow or Public header in
OPTIONS response (status code 200)
| http-robots.txt: 1 disallowed entry
|_/
|_http-title: SIAKAD STMIK AKBA : Login
81/tcp    open  tcpwrapped
|_auth-owners: ERROR: Script execution failed
(use -d to debug)
82/tcp    open  tcpwrapped
|_auth-owners: ERROR: Script execution failed
(use -d to debug)
.........................
```
**Fig. 4:** Showing open ports

Futhermore, after doing scanning as shown in Figure 4, the scanning of the IP target and scanning the results provide information about port and open service indicating that access to IP target is the real host server. In addition, the scanning results also give information that STMIK AKBA server is not placed in safe zone which is protected by firewall atau DMZ server. More details can be seen in Figure 5.

```
Starting Nmap 6.40 ( http://nmap.org ) at
2017-08-03 03:32 Malay Peninsula Standard Time
NSE: Loaded 110 scripts for scanning.
NSE: Script Pre-scanning.
Initiating Ping Scan at 03:32
Scanning siakad.akba.ac.id (               ) [4
ports]
Completed Ping Scan at 03:32, 0.29s elapsed (1
total hosts)
Initiating Parallel DNS resolution of 1 host.
at 03:32
Completed Parallel DNS resolution of 1 host.
at 03:33, 13.01s elapsed
```
**Fig. 5:** Target IP information

### 3.2. The results of analysis and application of new system

In a new system, the first step is istalling microtic until the selection of packages because it is needed to make a DMZ Zone for DMZ server as security system. Therefore, the importint thing to do first is installing microtic operation system in a computer server. When the istallation of microtic is done, the next is setting the IP in DMZ server that needs minimum 2 interfaces. Here, the first interface is for accessing internet and the second interface is as connector to DMZ zone where the STMIK AKBA server placed. For details of IP and interface of this research used DMZ, it can be seen in Table 1.

**Tabel 1:**
List of IP and Interface DMZ Server

| No. | Name of Interface | IP Address | Function |
|---|---|---|---|
| 1. | ether1 | 192.168.56.2 | Accessing to internet |
| 2. | ether2 | 192.168.0.1 | Accessing to DMZ Zone |

Based on Table 1, IP addresses are different, between ether 1 and ether 2. Interface for gateway accessing to internet in DMZ server is 192.168.56.1 and for web server simulation use IP 192.168.0.50 which is then put into DMZ server by typing the command as shown in Figure 6.

```
[admin@MikroTik]> ip address add ad-
dress=192.168.56.2/24
interface=ether1
[admin@MikroTik]> ip address add ad-
dress=192.168.0.1/24
interface=ether2
[admin@MikroTik]> ip route add gate-
way=192.168.56.1
```
**Fig. 6:** Command in DMZ Server

Then, IP checking was conducted to know the IP address that had been previously proggrammed by typing ip address print and ip route print as shown in Figure 7 below:

```
[admin@MikroTik]> ip address print
Flags: X – disabled, I – invalid, D – dynamic
#   ADDRESS            NETWORK          INTERFACE
0   192.168.56.2/24    192.168.56.0     ether1
1   192.168.0.1/24     192.168.0.0      ether2
......................
[admin@MikroTik]> ip route print
Flags: X – disabled, A – active, D – dynamic
C – connect, S – static, r – rip, b – bgp, o –
ospf, m – mme, B – blackhole, U – unreachable,
P – prohibit
#       DST-ADDRESS        PREF-SRC
0 A S   0.0.0.0/0
1 ADC   192.168.0.0/24     192.168.0.1
2 ADC   192.168.56.0/24    192.168.56.2

GATEWAY           DISTANCE
192.168.56.1          1
ether2                0
ether1                0
```
**Fig. 7:** IP Address dan IP Route

It should be noted that microtic supports the firewall in DMZ so testing and setting in STMIK AKBA server are not as complicated as we think. Therefore, in setting the DMZ parameter in microtic, it needs some commands as shown in Figure 8.

```
/ip firewall filter
add chain=forward connection-state=established
comment="allow established connections"
add chain=forward connection-state=related
comment="allow related connections"
add chain=forward connection-state=invalid
action=drop comment="drop invalid connections"
```
**Fig. 8:** Connection settings

The functions of commands in Figure 8 are to minimize and mantain the valid connection and block invalid connection. It can be proven by showing and releasing information about open port service in server of siakad.akba.ac.id where the port 80 for web server service http, port 225 for SSH service and port 443 for web server service https and others ports are closed. The scanning results can be seen in Figure 9.

Of three ports shown in Figure 9, all are open because they have been allowed by DMZ server to be accessed from outer network or internet. Yet, other ports are closed because they are not allowed by DMZ server as shown in port 22. Futhermore, Figure 10 explains the testing results which inform that scanning tool done can not detect the host location of siakda.akba.ac.id clearly because the access from outer network or internet is not directly connected to server of siakda.akab.ac.id but it is protected by DMZ server. Therfore, host that is likely accessed from outside network or internet is host DMZ server.

```
Nmap scan report for siakad.akba.ac.id
(            )
Host is up (0.38s latency).
Not shown: 991 filtered ports
PORT     STATE  SERVICE          VERSION
22/tcp   closed ssh
80/tcp   open   http             Apache httpd
|_http-methods: No Allow or Public header in
OPTIONS response (status code 301)
|_http-title: Did not follow redirect to
https://siakad.akba.ac.id/
255/tcp  open   ssh              (protocol
2.0)
| ssh-hostkey: 1024
c1:8d:fe:41:3c:b7:e0:93:a3:9f:2f:b2:8b:94:0c:0
0 (DSA)
| 2048
f2:df:f5:cf:ba:ef:1a:e3:22:0f:fe:73:d3:a4:c2:1
c (RSA)
|_256
47:a9:7d:94:4d:3b:1f:69:8c:db:89:22:b9:ee:38:6
a (ECDSA)
256/tcp  closed fw1-secureremote
443/tcp  open   ssl/https?
|_http-methods: No Allow or Public header in
OPTIONS
```

**Fig. 9:** Application result of DeMilitarized Zone

```
Aggressive OS guesses: Fortinet FortiGate-50B
or 310B firewall (89%), Fortinet FortiGate-60B
or -100A firewall (89%), Vonage V-Portal VoIP
adapter (89%), Cisco Unified Communications
Manager VoIP adapter (88%), Linksys WRV200
wireless broadband router (88%), DD-WRT v23
(Linux 2.4.36) (88%), DD-WRT v24-sp2 (Linux
2.4.36) (88%), Vyatta router (Linux 2.6.26)
(88%), Linux 2.6.18 (88%), Linux 2.6.22
(Kubuntu, x86) (88%)
No exact OS matches for host (test conditions
non-ideal).
```

**Fig. 10:** Scanning host server

In additin, the testing results by using DDOS (Distributed Daniel Of Services) method or Brute Force with Loic tool also provide information that cracker could not detect the location of the real host server. For details, it can be seen in Figure 12. Here, in the menu of select your target, URL or IP can be input from server target that will be attacked, then the targetted of port content by selecting the methods and the number of threats to do attacks, then klik *IMMA CHARGIN MAH LAZER*.

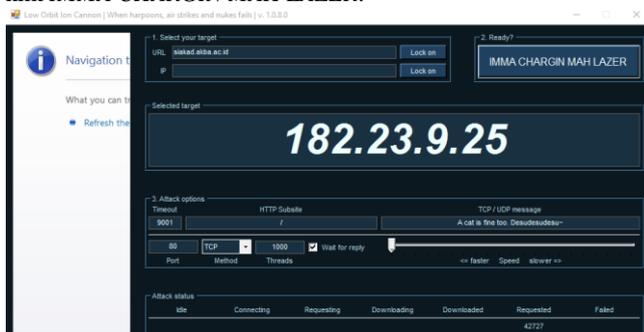

**Fig. 11:** Attack to IP server of STMIK AKBA

When having been attacked, DMZ server automatically will detect that attacks by putting them into data base of microtic address list. After microtic detects the the existing of DDOS or brute force attacks, IP automatically does reaccess server because it has been blocked by microtic. The bloking results can be seen in Figure 12.

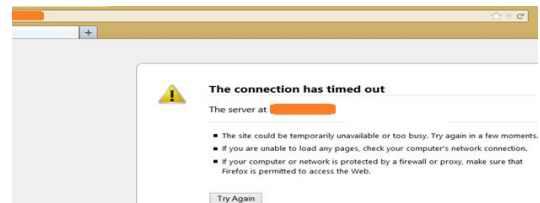

**Fig. 12:** Unable accessed Web server

The testing results are strenthened by the research results of [14] which revealed that applying Militarized Zone will provide better security both for infrastuctures of wired and wireless network. It is stated by [10] that DMZ is a major concept in hardware firewall which functions to improve the server security of web testing system especially internal and external network that are closedly related to the using of microtic device to control the network.

## 4. Conclusion

Based on the results of analysis and discussion that have been previously explained, it can be concluded that the application of DeMilitarized Zone method in microtic device in STMIK AKBA is running well which is shown on the output of scanning port results that is allowed by DMZ. Then, the scanning results by using DDOS (Distributed Daniel Of Service) technique also shows that the siakad server of STMIK AKBA is safe and inaccessible by intruders because DMZ server automatically detect the existence of IP access.

## Acknowledgement

Our highest gratitude to the head of STMIK AKBA and ICT head who had allowed us to conduct research in STMIK AKBA Laboratory.

## References

[1] S. Dandamudi and T. Eltaeib, "Firewalls Implementation in Computer Networks and Their Role in Network Security," *J. Multidiscip. Eng. Sci. Technol.*, vol. 2, no. 3, pp. 408–411, 2015.
[2] B. Souley and H. Abubakar, "A CAPTCHA – BASED INTRUSION DETECTION MODEL," *Int. J. Softw. Eng. Appl.*, vol. 9, no. 1, pp. 29–40, 2018.
[3] M. A. Al-ghazal and M. J. Aljubran, "Upstream Operations : Cybersecurity and Generation Y," *Saudi Aramco J. Technol.*, 2017.
[4] S. Prabhakar, "NETWORK SECURITY IN DIGITALIZATION : ATTACKS AND DEFENCE," *Int. J. Res. Comput. Appl. Robot.*, vol. 5, no. 5, pp. 46–52, 2017.
[5] A. A. A. Hadi, "Performance Analysis of Big Data Intrusion Detection System over Random Forest Algorithm," *Int. J. Appl. Eng. Res.*, vol. 13, no. 2, pp. 1520–1527, 2018.
[6] A. Bala, "Security Attacks and Challenges of Wireless Sensor Network," *Int. J. Sci. Res. Comput. Sci. Eng. Inf. Technol.*, vol. 3, no. 1, pp. 765–770, 2018.
[7] B. Rababah, S. Zhou, and M. Bader, "Evaluation the Performance of DMZ," *Assoc. Mod. Educ. Comput. Sci.*, no. January, pp. 0–13, 2018.
[8] S. M. Hashemi, J. He, and A. E. Basabi, "Multi-objective Optimization for Computer Security and Privacy," *Int. J. Netw. Secur.*, vol. 19, no. 3, pp. 394–405, 2017.
[9] M. KHAN, "Computer security in the human life," *Int. J. Comput. Sci. Eng.*, vol. 6, no. November, 2017.
[10] V. Selvi, R. Sankar, and R. Umarani, "The Design and Implementation of On-Line Examination Using Firewall security," *IOSR J. Comput. Eng.*, vol. 16, no. 6, pp. 20–24, 2014.
[11] M. Zammani and R. Razali, "An Empirical Study of Information Security Management Success Factors," *Int. J. Adv. Sci. Eng. Inf. Technol.*, vol. 6, no. 6, pp. 904–913, 2016.
[12] H. Liu, "Design and Implementation of Computer Network Vulnerability Assessment System," in *5th International Conference on Computer, Automation and Power Electronics*, 2017, pp. 145–149.

[13] M. Behi, M. Ghasemigol, and H. Vahdat-nejad, "A New Approach to Quantify Network Security by Ranking of Security Metrics and Considering Their Relationships," *Int. J. Netw. Secur.*, vol. 20, no. 1, pp. 141–148, 2018.

[14] S. Shrimali, "DeMilitarized Zone : Network Architecture for Information Security," *Int. J. Comput. Appl.*, vol. 174, no. 5, pp. 16–19, 2017.

[15] Mansyur and Muliana, "Detecting Differential Item Functioning and Differential Test Functioning on Math School Final-exam," *Int. J. Adv. Sci. Eng. Inf. Technol.*, vol. 6, no. 4, pp. 437–440, 2016.

[16] A. Iskandar, "The Effect of Open Book Test Model in Improving Students ' Learning Motivation," in *2nd International Conference on Education, Science, and Technology*, 2017, vol. 149, pp. 204–206.

[17] A. S. Ahmar *et al.*, "Lecturers' Understanding on Indexing Databases of SINTA, DOAJ, Google Scholar, SCOPUS, and Web of Science: A Study of Indonesians," *J. Phys. Conf. Ser.*, vol. 954, 2018.

[18] L. Akbay and M. Kaplan, "Transition to multidimensional and cognitive diagnosis adaptive testing: an overview of cat," *Online J. New Horizons Educ.*, vol. 7, no. 1, pp. 206–214, 2017.

[19] M. Rezaie and M. Golshan, "Computer Adaptive Test (CAT): Advantages and Limitations," *Int. J. Educ. Investig.*, vol. 2, no. 5, pp. 128–137, 2015.

[20] A. Puri and N. Sharma, "A Novel Technique For Intrusion Detection System For Network Security Using Hybrid SVM-CART," *Int. J. Eng. Dev. Res.*, vol. 5, no. 2, pp. 155–161, 2017.